\documentclass[reprint, amsmath, amssymb, superscriptaddress, prb]{revtex4-2}
\usepackage{graphicx}
\usepackage{hyperref}
\usepackage{xcolor}
\usepackage{bbm}
\usepackage{stmaryrd}

\begin{document}
\title{The scattering coefficients of superconducting microwave resonators:\\
I. Transfer-matrix approach} 

\author{Qi-Ming Chen}
\email{qiming.chen@wmi.badw.de}
\affiliation{Walther-Mei{\ss}ner-Institut, Bayerische Akademie der Wissenschaften, 85748 Garching, Germany}
\affiliation{Physik-Department, Technische Universit{\"a}t M{\"u}nchen, 85748 Garching, Germany}

\author{Meike Pfeiffer}
\affiliation{Physik-Department, Technische Universit{\"a}t M{\"u}nchen, 85748 Garching, Germany}

\author{Matti Partanen}
\affiliation{Walther-Mei{\ss}ner-Institut, Bayerische Akademie der Wissenschaften, 85748 Garching, Germany}

\author{Florian Fesquet}
\author{Kedar E. Honasoge}
\author{Fabian Kronowetter}
\author{Yuki Nojiri}
\author{Michael Renger}
\author{Kirill G. Fedorov}
\affiliation{Walther-Mei{\ss}ner-Institut, Bayerische Akademie der Wissenschaften, 85748 Garching, Germany}
\affiliation{Physik-Department, Technische Universit{\"a}t M{\"u}nchen, 85748 Garching, Germany}

\author{Achim Marx}
\affiliation{Walther-Mei{\ss}ner-Institut, Bayerische Akademie der Wissenschaften, 85748 Garching, Germany}

\author{Frank Deppe}
\email{frank.deppe@wmi.badw.de}
\author{Rudolf Gross}
\email{rudolf.gross@wmi.badw.de}
\affiliation{Walther-Mei{\ss}ner-Institut, Bayerische Akademie der Wissenschaften, 85748 Garching, Germany}
\affiliation{Physik-Department, Technische Universit{\"a}t M{\"u}nchen, 85748 Garching, Germany}
\affiliation{Munich Center for Quantum Science and Technology (MCQST), Schellingstr. 4, 80799 Munich, Germany}

\date{\today}

\begin{abstract}
	We describe a unified classical approach for analyzing the scattering coefficients of superconducting microwave resonators with a variety of geometries. To fill the gap between experiment and theory, we also consider the influences of small circuit asymmetry and the finite length of the feedlines, and describe a procedure to correct them in typical measurement results. We show that, similar to the transmission coefficient of a hanger-type resonator, the reflection coefficient of a necklace- or bridge-type resonator does also contain a reference point which can be used to characterize the electrical properties of a microwave resonator in a single step. Our results provide a comprehensive understanding of superconducting microwave resonators from the design concepts to the characterization details.
\end{abstract}

\maketitle

\section{Introduction}
Superconducting microwave resonators are indispensable components in superconducting quantum circuits \cite{Gu2017}. Owing to the high flexibility of circuit design, the resonators can be made and coupled to other components, or an external circuitry, in various ways with different emphases \cite{Schoelkopf2008}. For example, a necklace-type resonator, where the feedline(s) and the resonator are coupled end-to-end, is often used to control and couple different qubits \cite{Schuster2005, Wallraff2005, Majer2007, Sillanpaeae2007, Hofheinz2008, Hofheinz2009, DiCarlo2009, DiCarlo2010, Chow2011, Chow2012, Yin2013, Sete2013, Goetz2017, Goetz2018}, while a hanger-type resonator, where one end of the bare resonator is coupled to either side of the feedline, is more common for reading out the quantum information \cite{Barends2013, Chen2014, Jeffrey2014, Song2017, Song2019, Wang2020, Guo2021, Ma2019, Mundada2019, Andersen2020, Lacroix2020, Besse2020}. Depending on the detailed geometry of the circuit designs, the scattering coefficients of superconducting microwave resonators may show fundamentally different line shapes that carry the information of different physical processes. However, relevant discussions are either focused on the hanger-type $\lambda/4$ resonators or restricted to the transmission coefficient solely \cite{Gao2007, O'Connell2008, Song2009, Song2009a, Gao2011, Sage2011, Megrant2012, Geerlings2012, Bruno2015, Goetz2016}. A comprehensive study of the scattering coefficients, especially on the reflection coefficient of necklace-type resonators, is still missing in the literature. As a consequence, it is well known that a hanger-type $\lambda/4$ resonator can be characterized by measuring only the transmission coefficient, which contains a reference point that distinguishes the internal and coupling quality factors from the line shape \cite{Petersan1998, Khalil2012, Deng2013, Probst2014, McRae2020}. However, to date one still has to combine transmission measurements with either detailed cable calibration or finite-element simulations to characterize a necklace-type resonator \cite{Frunzio2004, Goeppl2008, Yeh2013, Ranzani2013, Cataldo2015, Wang2019}. This complexity limits not only the reliability of the characterization results but also the applicability of the necklace-type resonators to a complex circuit. 

Here, and also in a parallel paper \cite{Chen2021b}, we study the scattering coefficients of superconducting microwave resonators in either classical or quantum perspectives. In this work, we employ the transfer-matrix method in microwave engineering and derive the analytical descriptions of the scattering coefficients for a general resonator \cite{Pozar2011}. We also explore the physical origin of the line-shape distortions that are commonly seen in experiments, and describe a procedure to remove these effects. Finally, we experimentally demonstrate that the reflection coefficient of a necklace-type resonator does contain a reference point, which can be used to correct the experimental imperfections and characterize the electric properties of the resonator. These results provide a systematic study of the scattering coefficients of superconducting microwave resonators in the classical perspective. 

The rest of this paper is organized as follows: In Sec.\,\ref{sec:unload}, we introduce the circuit diagrams of different types of superconducting microwave resonators, and review their electrical properties such as resonant frequency and quality factors. Next, we outline the transfer-matrix method and derive the ideal scattering coefficients of different resonators in Sec.\,\ref{sec:load}. We study how the circuit asymmetries and the finite length of the feedlines can influence the measured scattering coefficients in Sec.\,\ref{sec:distortions}, and obtain a general description of the scattering response. We demonstrate these results in an experiment in Sec.\,\ref{sec:experiment}, and, finally, we conclude this study in Sec.\,\ref{sec:conclusions}. Detailed derivations of the scattering coefficients for different types of resonators can be found in Appendices\,\ref{app:necklace2}-\ref{app:asy_cross2}. The detailed procedure for correcting the experimental imperfections in measured scattering coefficients is described in Appendix\,\ref{app:circle-fit}

\section{Circuit description of microwave resonators} \label{sec:unload} 
\begin{figure}
\centering
\includegraphics[width=\columnwidth]{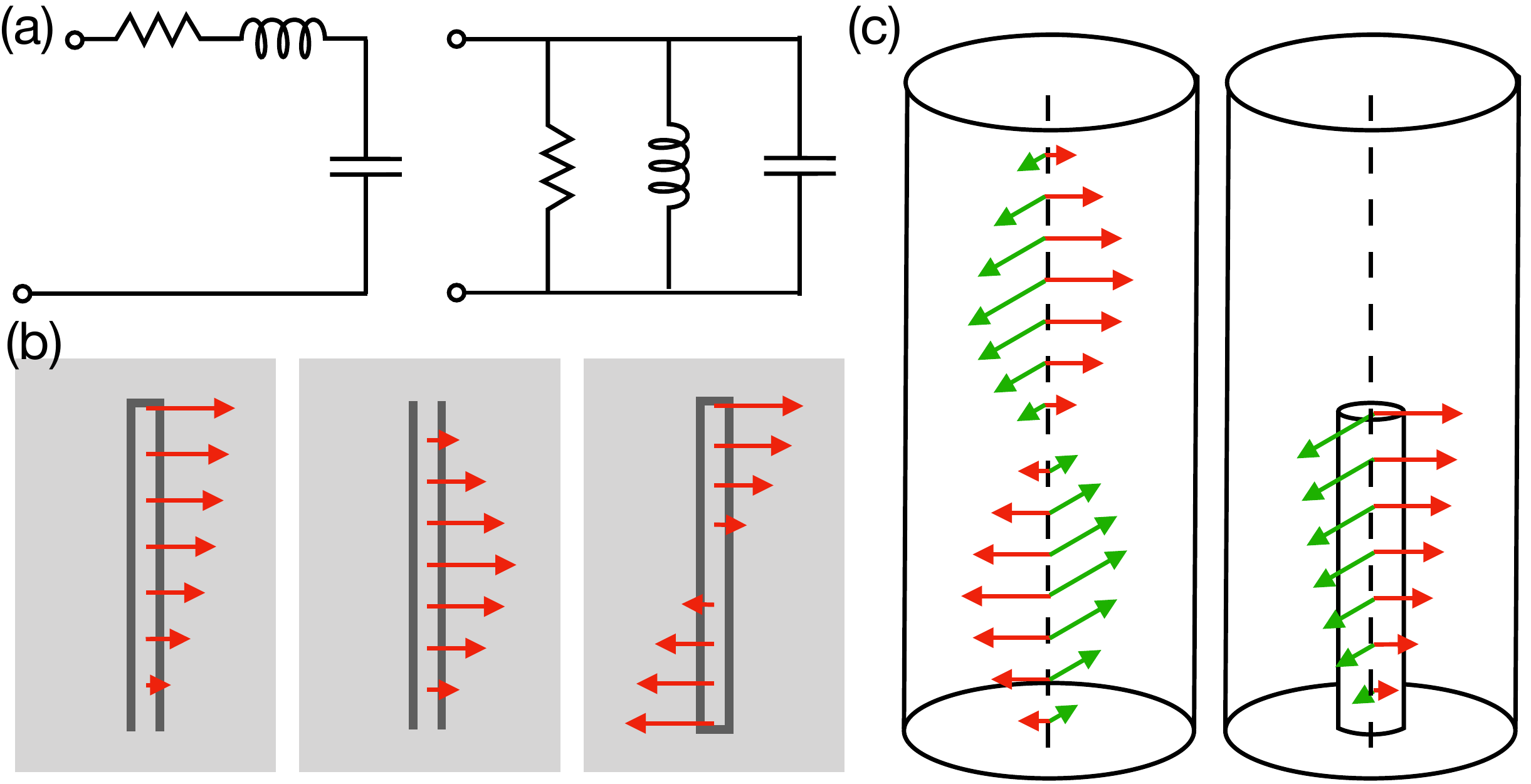}
\caption{Schematics of several different microwave resonators. (a) In lumped-element circuits, the series (left) and parallel (right) RLC resonant circuits are the two fundamental types of microwave resonators. (b) In distributed-element circuits, there are three types of transmission line resonators: the short-circuited $\lambda/4$ (left), the short-circuited $\lambda/2$ (middle), and the open-circuited $\lambda/2$ (right) resonators. (c) Microwave resonators  can also be made by bulky 3D devices, which can be modeled by a short-circuited $\lambda/2$ (left) or a short-circuited $\lambda/4$ (right) transmission line resonator. In all the panels, red and green vectors depict the special modes of the resonator, which are determined by the boundary conditions.}
\label{fig:unloaded}
\end{figure}

In lumped-element circuits, there are two fundamental types of circuit diagrams that can be modeled as a \textit{zero}-dimensional microwave resonator. As schematically shown in Fig.\,\ref{fig:unloaded}(a), the first one is called the series RLC resonator, which has an input impedance of 
\begin{equation}
	Z_{\rm s} = R + j\omega L + \frac{1}{j\omega C}. \label{eq:series}
\end{equation}
Here, $R$, $L$, and $C$ are the resistance, inductance, and capacitance of the circuit, $j$ is the imaginary unit that follows the convention of electrical engineering. The resonance occurs at ${\rm Im}\left(Z_{\rm s}\right)=0$, which corresponds to a resonant frequency $\omega_0=1/\sqrt{LC}$ and a (internal) quality factor $Q_{\rm i} = \omega_0L/R = 1/\left(\omega_0 RC\right)$ \cite{Pozar2011}. The second one, as shown in Fig.\,\ref{fig:unloaded}(b), is called the parallel RLC resonator with an input impedance of
\begin{equation}
	Z_{\rm p} = \left( \frac{1}{R} + \frac{1}{j\omega L} + j\omega C \right)^{-1}. \label{eq:parallel}
\end{equation}
Similarly, one can calculate the resonance frequency and the (internal) quality factor as $\omega_0=1/\sqrt{LC}$ and $Q_{\rm i} = R/\left(\omega_0L\right) = \omega_0 RC$, respectively \cite{Pozar2011}. 

In comparison, a finite length of transmission line with proper boundary conditions can also be described as a microwave resonator in one dimensional. Depending on the load impedance $Z_{\rm L}$ and the length $l$, the input impedance of a transmission line, when looking towards one end, can be written as \cite{Pozar2011}
\begin{equation}
	Z(l) = Z_0 \frac{Z_{\rm L} + Z_0 \tanh \gamma l}{Z_0 + Z_{\rm L}\tanh \gamma l}.
\end{equation}
Here, $\gamma = \alpha + j\beta$ is the complex propagation constant of the microwave field, $Z_0$ is the characteristic impedance of the transmission line. By assuming a small damping rate of the transmission line, i.e., $\alpha l \ll 1$, and confining our discussion in a small frequency range around the resonant frequency, i.e., $\left|\Delta\right| \ll \omega_0$ with $\Delta = \omega - \omega_{0}$, one can get three types of microwave resonators as shown in Fig.\,\ref{fig:unloaded}(b). They are (i) the short-circuited $\lambda/4$ resonator with  
\begin{equation}
	Z_{\lambda/4} = \frac{Z_0}{\alpha l + j\pi \Delta/2\omega_0}\,\text{and}\,
	\omega_0=\frac{\pi v_{\rm p}}{2l},
\end{equation}
(ii) the short-circuited $\lambda/2$ resonator with
\begin{equation}
	Z_{\lambda/2}^{\rm (short)} = Z_0 \left( \alpha l + j\pi\Delta/\omega_0 \right)\,\text{and}\,
	\omega_0=\frac{\pi v_{\rm p}}{l},
\end{equation}
and (iii) the open-circuited $\lambda/2$ resonator with
\begin{equation}
	Z_{\lambda/2}^{\rm (open)} = \frac{Z_0}{\alpha l + j\pi \Delta/\omega_0}\,\text{and}\,
	\omega_0=\frac{\pi v_{\rm p}}{l}.
\end{equation}
Here, $v_{\rm p}=\omega/\beta$ is the phase velocity of the propagating microwave field in the transmission line. 

Comparing these results with the two lumped-element resonators, we observe that the short-circuited $\lambda/4$ resonator and the open-circuited $\lambda/2$ resonator are equivalent to a parallel RLC resonator with $R=Z_0/(\alpha l)$, $L=1/\left(\omega_{0}^2 C\right)$, and $C=\pi /(4\omega_{0}Z_0)$ or $C=\pi /(2\omega_{0}Z_0)$, respectively. The short-circuited $\lambda/2$ resonator is equivalent to a series RLC resonator with $R=Z_0\alpha l$, $L=\pi Z_0/(2\omega_{0})$, and $C=1/(\omega_{0}^2 L)$. However, the (internal) quality factor has the same definition for all the three resonators: $Q_{i}=\beta/(2\alpha)$, where $\beta = 2\pi/\lambda$. 

The above discussions also apply to 3D microwave resonators as shown in Fig.\,\ref{fig:unloaded}(c), which attract an increasing amount of interests during the past decades for their superior quality factors. The inner surface of a 3D resonator naturally defines the nodes of the spacial modes of the electrical field, while the anti-nodes are located either at the anti-nodes of the standing waves inside the cavity \cite{Paik2011, Rigetti2012, Brecht2015, Brecht2016, Brecht2017, Xie2018, Abdurakhimov2019, Romanenko2014, Romanenko2020, Lei2020, Chakram2021}, or at the top of a $\lambda/4$-long waveguide lead standing inside the inner space \cite{Reagor2013, Reagor2016, Kudra2020, Wang2021b}. If the electrical fields can be fairly described as one-dimensional functions of the coordinator, the two types of 3D resonators can be equivalently described by a shorted-circuited $\lambda/2$ or a shorted-circuited $\lambda/4$ transmission line resonator, respectively. In this regard, we do not distinguish coplanar waveguide resonators and 3D resonators in the current discussion. A careful distinction may be necessary when studying exotic resonator designs, for example, the 2D resonators introduced in Refs.\,\cite{Minev2013, Minev2016}.

\section{Ideal scattering coefficients of microwave resonators} \label{sec:load}
\begin{figure}
\centering
\includegraphics[width=\columnwidth]{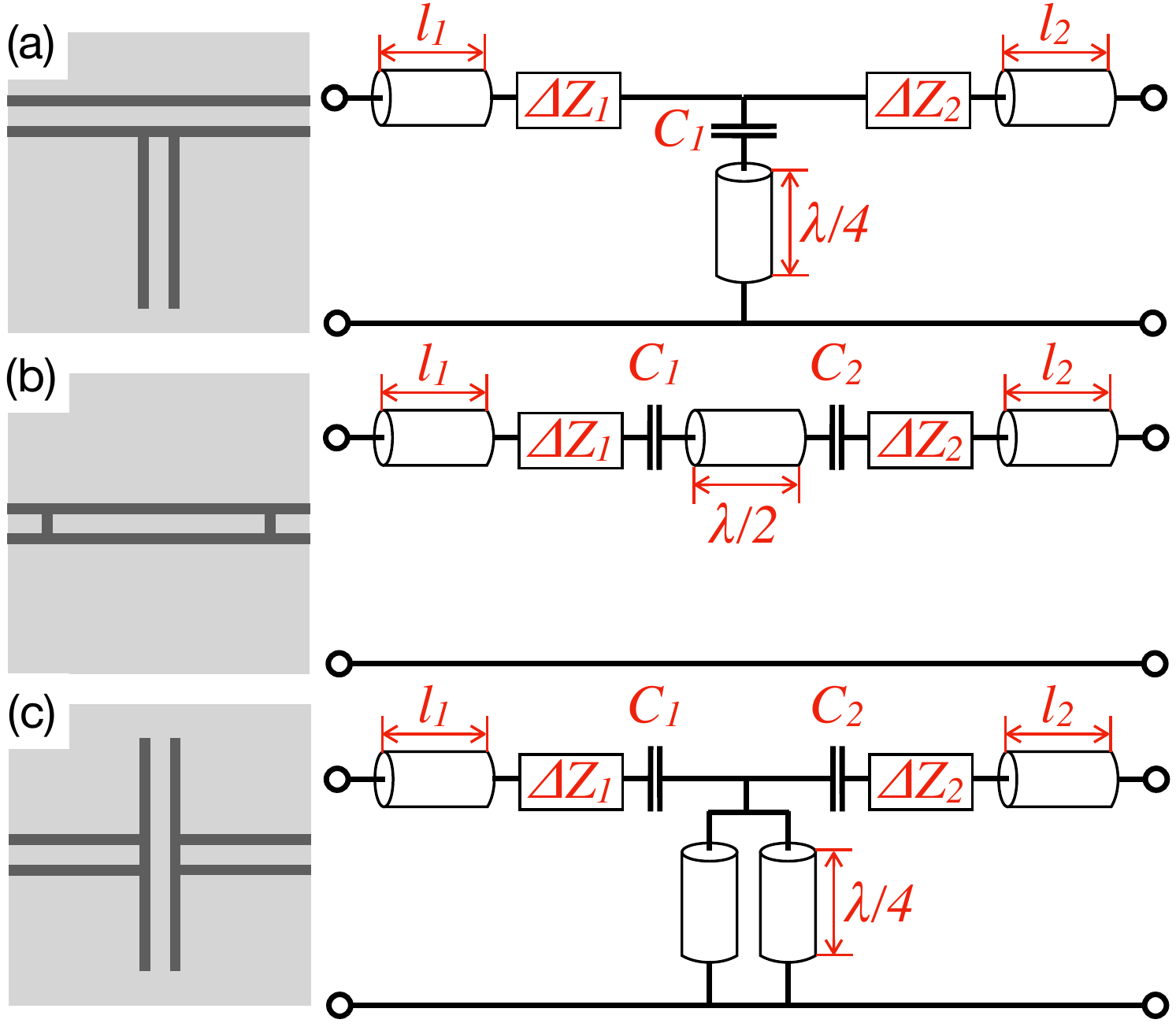}
\caption{Schematics of three typical resonator designs and the corresponding circuit diagrams. They are (a) the hanger-type $\lambda/4$ resonator, (b) the necklace-type $\lambda/2$ resonator, and (c) the bridge-type $\lambda/2$ resonator. In all the panels, we denote $l_1$ and $l_2$ as the lengths of the feedlines that couple to the resonator, $C_1$ and $C_2$ the coupling capacitors, and $\Delta{Z_1}$ and $\Delta{Z_2}$ the asymmetries that may exist in the circuit. Without loss of generality, we label the left and right ports by $1$ and $2$.}
\label{fig:load}
\end{figure}

To measure the electrical properties of a microwave resonator, such as the resonant frequency and the quality factor, one has to couple it to an external circuitry, called the load, and measure the scattering coefficients, as schematically shown in Fig.\,\ref{fig:load}(a)-(c). However, the coupling also leads to an inevitable change of the electric properties to be measured. By convention, we define the loaded quality factor, $Q_{\rm l}$, as a combination of two terms \cite{Pozar2011}
\begin{equation}
	\frac{1}{Q_{\rm l}} = \frac{1}{Q_{\rm i}} + \frac{1}{Q_{\rm c}}.
\end{equation}
Here, $Q_{\rm c}$ is defined as the coupling quality factor which describes the power-loss ratio induced by the external circuity, $Q_{\rm i}$ is the internal quality factor that characterizes the bare resonator loss. In this section, we study the ideal scattering coefficients of three types of resonators that are commonly seen in superconducting quantum circuits. We assume \textit{zero}-length transmission feedlines, i.e., $l_1,l_2=0$ and \textit{zero} circuit asymmetries, $\Delta{Z_1},\Delta{Z_2}=0$ for now for simplicity.

\subsection{The hanger-type $\lambda/4$ resonator}\label{sec:loaded_hanger4}
The hanger-type $\lambda/4$ resonator is a short-circuited transmission line with one end shorted to ground and the other side-coupled to a one-dimensional waveguide. The schematic of this resonator is shown in Fig.\,\ref{fig:load}(a), where we define the coupling capacitance as $C_{1}$, and label the left and right ports as port $1$ and $2$, respectively. The elements of the transfer matrix read $A=D=1$, $B=0$, and $C=1/Z$ with $Z=1/j\omega C_{1} + Z_{\lambda/4}$. Following the standard procedure, the scattering coefficients can be readily calculated with the following relations \cite{Pozar2011}
\begin{align}
	S_{11} &= \frac{A+B/Z_0-CZ_0-D}{A+B/Z_0+CZ_0+D}, \label{eq:s_11}\\
	S_{12} &= \frac{2(AD-BC)}{A+B/Z_0+CZ_0+D}, \label{eq:s_12}\\
	S_{21} &= \frac{2}{A+B/Z_0+CZ_0+D}, \label{eq:s_21}\\
	S_{22} &= \frac{-A+B/Z_0-CZ_0+D}{A+B/Z_0+CZ_0+D}. \label{eq:s_22}
\end{align}

To simplify the expressions, let us consider first a lossless resonator. Knowing that the resonance occurs at ${\rm Im}\left( Z \right)=0$, we can calculate the resonant frequency as
\begin{equation}
	\omega_{\rm r} \approx \omega_0 - \frac{2Z_0 C_{1} \omega_0^2}{\pi}.
\end{equation}
This result still holds for lossy resonators as long as $Q_{\rm i} \gg \sqrt{Q_{\rm c}}$, where 
\begin{equation}
	Q_{\rm c} = \frac{\pi}{2\omega_{\rm r}^2Z_0^2 C_{1}^2}. \label{eq:qc_hanger_4}
\end{equation}
This relation is valid for typical experimental situations in superconducting quantum circuits. In this regard, we rewrite the input impedance of the vertical branch close to the resonant frequency as $Z \approx Z_0 Q_{\rm c}\left[ 1/\left(2Q_{\rm i}\right)+j\delta \right]$ with $\delta =\left( \omega - \omega_{\rm r} \right)/\omega_{\rm r}$. Inserting this result in Eqs.\,\eqref{eq:s_11}-\eqref{eq:s_22}, we obtain the scattering coefficients as
\begin{align}
	S_{11} = S_{22} \approx& -\frac{Q_{\rm l}/Q_{\rm c}}{1+j2Q_{\rm l}\delta}, \label{eq:hanger4_s11}\\
	S_{21} = S_{12} \approx& 1 -\frac{Q_{\rm l}/Q_{\rm c}}{1+j2Q_{\rm l}\delta}. \label{eq:hanger4_s21}
\end{align}

The physical meaning of the parameter, $Q_{\rm c}$, can be understood in the perspective of the Norton's equivalent lumped-element circuit \cite{Mazin2004, Gao2008, Jerger2013, Geerlings2013, Reagor2015}. Assuming that $\omega_0 C_{1}Z_0 \ll 1$, the loaded quality factor can be written as
\begin{equation}
	Q_{\rm l} = \omega_{\rm r} C \left( \frac{1}{R} + \frac{\omega_{\rm r}^2 Z_0 C_{1}^2}{2} \right)^{-1}. \label{eq:qc4}
\end{equation}
For $\lambda/4$ resonators, we have $\omega_{\rm r} C \approx \pi/4Z_0$, such that the coupling quality factor is $Q_{\rm c}=\pi/\left(2Z_0^2\omega_r^2 C_{1}^2\right)$. Thus, the parameter $Q_{\rm c}$ defined in Eq.\,\eqref{eq:qc_hanger_4} can be interpreted as the coupling quality factor of a hanger-type $\lambda/4$ resonator, as is indicated in the notation. 

The above discussion can also be generalized to a hanger-type $\lambda/2$ resonator. If one neglects the coupling between the open-end of the resonator and the ground plane, the scattering coefficients of a hanger-type $\lambda/2$ resonator are exactly the same with those of a hanger-type $\lambda/4$ resonator, as shown in Eqs.\,\eqref{eq:hanger4_s11}-\eqref{eq:hanger4_s21}. However, the resonant frequency, $\omega_{\rm r} \approx \omega_0 - Z_0 C_{1} \omega_0^2/\pi$, and the coupling quality factor, $Q_{\rm c} = \pi/\left(\omega_{\rm r}^2Z_0^2 C_{1}^2\right)$, are defined differently from those of an $\lambda/4$ resonator. 

As a crosscheck of the above results, we compare in Fig.\,\ref{fig:simulation}(a) the analytical formulae with the numerically simulated scattering coefficients of a hanger-type $\lambda/4$ resonator. Here, the parameters are chosen such that $\omega_{0}=2\pi \times 6.75\,{\rm GHz}$, $Q_{\rm i}=31416$ for a bare resonator. In the presence of the coupling, the analytical expressions in Eqs.\,\ref{eq:hanger4_s11}-\eqref{eq:hanger4_s21} predict the resonant frequency, $\omega_{\rm r}=2\pi \times 6.659\,{\rm GHz}$, and the quality factors, $Q_{\rm c}=3589$, $Q_{\rm l}=3221$. These values exhibit an excellent fit to the numerical simulation results.

\subsection{The necklace-type $\lambda/2$ resonator}\label{sec:loaded_necklace2}
The necklace-type resonator, as schematically shown in Fig.\,\ref{fig:load}(b), consists of an open-circuited transmission line which is capacitively coupled to two feedlines through the capacitors, $C_{1}$ and $C_{2}$, at the two ends, respectively. The elements of the transfer matrix of the coupling capacitors read $A=D=1$, $B=1/j\omega C_1$ or $1/j\omega C_2$, and $C=0$. They are $A=D=\cosh \gamma l$, $B=Z_0\sinh \gamma l$, and $C=\left(1/Z_0\right)\sinh \gamma l$ for a bare open-circuited $\lambda/2$ resonator. Following a similar treatment as for the hanger-type resonators, we obtain the scattering coefficients as (see Appendix\,\ref{app:necklace2} for details)
\begin{align}
	S_{11} \approx& 1- \frac{2Q_{\rm l}/Q_{{\rm c},1}}{1 + j2Q_{\rm l}\delta}, \label{eq:necklace2_s11}\\
	S_{21} =& S_{12} \approx \frac{2Q_{\rm l}/\sqrt{Q_{{\rm c},1}Q_{{\rm c},2}}}{1 + j2Q_{\rm l}\delta}, \label{eq:necklace2_s21}\\
	S_{22} \approx& 1- \frac{2Q_{\rm l}/Q_{{\rm c},2}}{1 + j2Q_{\rm l}\delta}. \label{eq:necklace2_s22}
\end{align}
Here, $1/Q_{c} = 1/Q_{{\rm c},1} + 1/Q_{{\rm c},2}$ with 
\begin{equation}
	Q_{{\rm c},k} = \frac{\pi}{2\omega_r^2Z_0^2 C_k^2},~k=1,2.\label{eq:necklace_qc}
\end{equation}
The resonant frequency is 
\begin{equation}
	\omega_{\rm r} = \omega_0 - \frac{Z_0\left( C_1 + C_2 \right)\omega_0^2}{\pi}. \label{eq:necklace_wr}
\end{equation} 

The physical meaning of the parameter, $Q_{\rm c}$, can also be understood in the perspective of the Norton's equivalent lumped-element circuit \cite{Schuster2007, Goeppl2009}. Assuming that $\omega_{0} C_{1} Z_0, \omega_{0} C_{2}Z_0 \ll 1$, the loaded quality factor reads
\begin{equation}
	Q_{\rm l} = \omega_{\rm r} C \left( \frac{1}{R} + \omega_{\rm r}^2 Z_0 C_{1}^2 + \omega_{\rm r}^2 Z_0 C_{2}^2 \right)^{-1},
\end{equation}
where $\omega_{\rm r} C = \pi/2Z_0$ for a $\lambda/2$ resonator. In this regard, the coupling quality factor can be written as
\begin{equation}
	\frac{1}{Q_{\rm c}} = \frac{2\omega_{\rm r}^2Z_0^2 C_1^2}{\pi} + \frac{2\omega_{\rm r}^2Z_0^2 C_2^2}{\pi},
\end{equation}
which is exactly the sum of the two parameters $1/Q_{{\rm c},1}$ and $1/Q_{{\rm c},2}$ defined in Eq.\,\eqref{eq:necklace_qc}.

The above discussion can also be generalized to a necklace-type $\lambda/4$ resonator, which is a single-port device with only the reflection coefficient shown in Eq.\,\eqref{eq:necklace2_s11}. The resonant frequency and the coupling quality factor are $\omega_{\rm r} \approx \omega_0 - 2Z_0 C_{1} \omega_0^2/\pi$ and $Q_{\rm c} = \pi/\left(4\omega_{\rm r}^2Z_0^2 C_{1}^2\right)$, respectively. We note that the transmission coefficient, $S_{21}$ in Eq.\,\eqref{eq:necklace2_s21}, is consistent with the results reported in the literature \cite{Frunzio2004, Goeppl2008}. However, the reflection coefficients, $S_{11}$ and $S_{22}$ in Eqs.\,\eqref{eq:necklace2_s11} and \eqref{eq:necklace2_s22}, have not been reported but have an important feature: The complex scattering coefficients form a circle which intersects with the real axis at a fixed point, $\left(1+j0\right)$, for $\delta \rightarrow \infty$. The circle radius, $r_{\rm c}$, equals to $Q_{\rm l}/\left(2Q_{\rm c}\right)$ for $C_{1}=C_{2}$. When $C_{1} \neq C_{2}$, the circle radii should be $Q_{\rm l}/Q_{{\rm c},1}$ and $Q_{\rm l}/Q_{{\rm c},2}$ for $S_{11}$ and $S_{22}$, respectively. These results are similar to the transmission coefficients of a hanger-type resonator.

\begin{figure}
  \centering
  \includegraphics[width=\columnwidth]{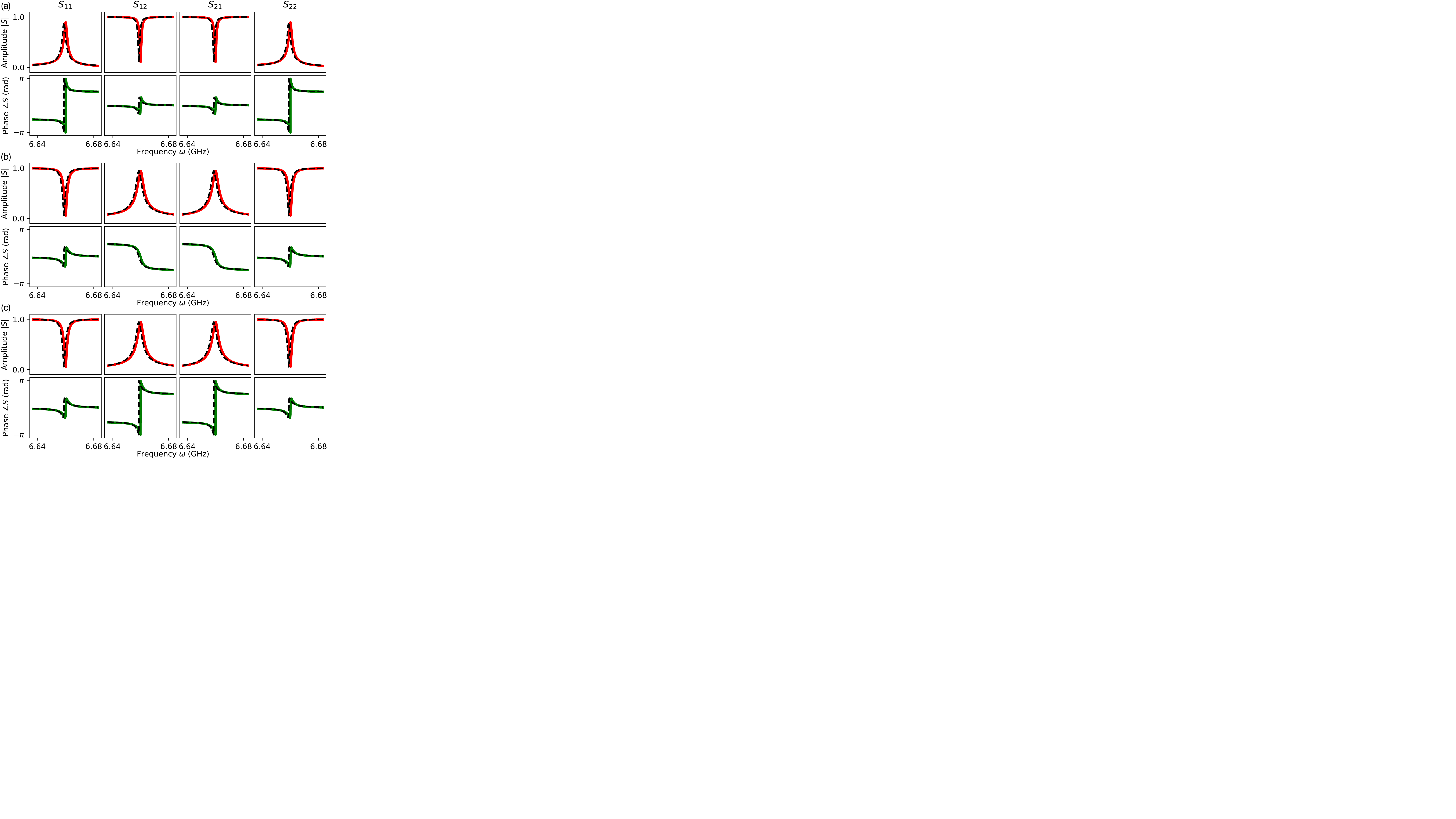}
  \caption{Simulated scattering coefficients of the three types of microwave resonators shown in Fig.\,\ref{fig:load}. They are (a) the hanger-type $\lambda/4$ resonator, (b) the necklace-type $\lambda/2$ resonator, and (c) the bridge-type $\lambda/2$ resonator. In all the panels, the  the black dashed curves represent the analytical results. They fit closely to the red and green curves, which are the numerical results calculated by using the transfer matrices.}
  \label{fig:simulation}
\end{figure}

We also compare the analytical formulae with the numerically simulated scattering coefficients of a necklace-type $\lambda/2$ resonator, as shown in Fig.\,\ref{fig:simulation}(b). Here, the parameters are set identical to Fig.\,\ref{fig:simulation}(a) except that the length of the transmission line is doubled. In the presence of a finite coupling, the analytical expressions in Eqs.\,\eqref{eq:necklace2_s11}-\eqref{eq:necklace2_s22} predict the resonant frequency, $\omega_{\rm r}= 2\pi\times 6.659\,{\rm GHz}$, and the quality factors $Q_{\rm c}=1795$, $Q_{\rm l}=1698$. These values are in excellent agreement with those obtain by the numerical simulation. We note that for the same $\lambda/2$ resonator coupled in either hanger- or necklace-type, the coupling quality factors can show a huge difference. For example, a hanger-type $\lambda/2$ resonator will lead to $Q_{\rm c}=7082$, which is $4$ times larger than that of a necklace-type resonator. In this regard, we may conclude that a necklace-type resonator is more suitable for driving a quantum system or reading out the quantum information, while a hanger-type resonator for storing quantum information. 

\subsection{The bridge-type $\lambda/2$ resonator} \label{sec:loaded_cross2}
There is a third type of superconducting microwave resonator which is rare in coplanar waveguide designs but common in 3D structures. The schematic is shown in Fig.\,\ref{fig:load}(c), where the two ends of the feedlines are coupled to the voltage anti-nodes of the bare resonator. Here, we restrict our discussion to the fundamental mode and figuratively name it a bridge-type $\lambda/2$ resonator. The circuit diagram is a combination of the hanger- and necklace-type resonators, which consists of two coupling capacitors, $C_1$ and $C_2$, and also two parallel short-circuited $\lambda/4$ resonators in the vertical branch. The elements of the transfer matrix read $A=1+2/\left(j\omega C_1 Z_{\lambda/4}\right)$, $B=-\left[j\omega \left(C_1+C_2\right)Z_{\lambda/4}+ 2\right]/\left(\omega^2 C_1 C_2 Z_{\lambda/4}\right)$, $C=2/Z_{\lambda/4}$, and $D=1+2/\left(j\omega C_2 Z_{\lambda/4}\right)$. The scattering coefficients can be readily obtained by following the standard procedure (see Appendix\,\ref{app:cross2} for details)
\begin{align}
	S_{11} \approx& 1- \frac{2Q_{\rm l}/Q_{{\rm c},1}}{1 + j2Q_{\rm l}\delta}, \label{eq:cross2_s11}\\
	S_{21} =& S_{12} \approx -\frac{2Q_{\rm l}/\sqrt{Q_{{\rm c},1}Q_{{\rm c},2}}}{1 + j2Q_{\rm l}\delta}, \label{eq:cross2_s21}\\
	S_{22} \approx& 1- \frac{2Q_{\rm l}/Q_{{\rm c},2}}{1 + j2Q_{\rm l}\delta}, \label{eq:cross2_s22}
\end{align}
where $1/Q_{\rm c} = 1/Q_{{\rm c},1} + 1/Q_{{\rm c},2}$ with 
\begin{equation}
	Q_{{\rm c},k} = \frac{\pi}{2\omega_r^2Z_0^2 C_k^2},~k=1,2. \label{eq:cross_qc}
\end{equation}
The resonance occurs at 
\begin{equation}
	\omega_{\rm r} = \omega_0 - \frac{Z_0\omega_0^2\left( C_1 + C_2 \right)}{\pi}. \label{eq:cross_wr}
\end{equation} 

Comparing with Eqs.\,\eqref{eq:necklace2_s11}-\eqref{eq:necklace2_s22}, the scattering coefficients of a bridge-type resonator have an almost identical form as for the necklace-type resonator except a $\pi$ phase difference in the transmission coefficients, $S_{12}$ and $S_{21}$. We also compare in Fig.\,\ref{fig:simulation}(c) the analytical formulae with the numerically simulated scattering coefficients of a bridge-type $\lambda/2$ resonator. Here, the parameters are set identical to Fig.\,\ref{fig:simulation}(b). In the presence of a finite coupling, the analytical expressions in Eqs.\,\eqref{eq:cross2_s11}-\eqref{eq:cross2_s21} predict the resonant frequency, $\omega_{\rm r}=2\pi\times 6.659\,{\rm GHz}$, and the quality factors $Q_{\rm c}=1795$, $Q_{\rm l}=1698$. These values are identical to those of a necklace-type $\lambda/2$ resonator.

\section{Practical distortions in scattering coefficients} \label{sec:distortions}
\subsection{Influence of small circuit asymmetry}
In contrast to the ideal circuits we described above, asymmetries often exist in real circuits in the form of mutual inductance \cite{Khalil2012}, impedance mismatch \cite{Deng2013}, etc. In this section, we keep the \textit{zero}-length feedline assumption but study the influence of a small circuit asymmetry on the scattering coefficients.

\subsubsection{Hanger-type resonators}
As is schematically shown in Fig.\,\ref{fig:load}(a), we consider small circuit asymmetries, $\Delta{Z_1}, \Delta{Z_2} \ll Z_0$, on both sides of the ideal system. The elements of the transfer matrix with asymmetry can be written as $A = 1 + \Delta{Z_1}/Z$, $B = \Delta{Z_1} + \Delta{Z_2} + \Delta{Z_1}\Delta{Z_2}/Z$, $C = 1/Z$, $D = 1+\Delta{Z_2}/Z$. Following the derivation in Appendix\,\ref{app:asy_hanger4}, we obtain the scattering coefficients as
\begin{align}
	S_{11} \approx& -\frac{Q_{\rm l}'/Q_{\rm c}'}{1+j2Q_{\rm l}\delta},\\
	S_{21} = S_{12} \approx& 1 -\frac{Q_{\rm l}'/Q_{\rm c}'}{1+j2Q_{\rm l}'\delta}, \\
	S_{22} \approx& -\frac{Q_{\rm l}'/Q_{\rm c}'}{1+j2Q_{\rm l}\delta}.
\end{align}
Because of the asymmetry, the loaded quality factor becomes $1/Q_{\rm l}' = 1/Q_{\rm i} + 1/Q_{\rm c}'$, where $Q_{\rm c}'=Q_{\rm c}Z_0\left( 1/Z_1 + 1/Z_2 \right)/2$ with $Z_1 = Z_0 + \Delta{Z_1}$ and $Z_2 = Z_0 + \Delta{Z_2}$. In this regard, both $Q_{\rm l}'$ and $Q_{\rm c}'$ can take complex values which indicate a loss mechanism different from a simple exponential energy decay. However, for small circuit asymmetries the real parts of $Z_1$ and $Z_2$ should still be dominant. We follow the convention and redefine the loaded quality factor as $1/Q_{\rm l}' = 1/Q_{\rm i} + {\rm Re}\left(1/Q_{\rm c}'\right)$ \cite{Khalil2012, Deng2013}. The information of the imaginary part, ${\rm Im}\left(1/Q_{\rm c}'\right)$, is interpreted as a small phase factor $\phi = -\arctan \left[ {\rm Im}\left( Q_{\rm c}'\right)/{\rm Re} \left(Q_{\rm c}' \right) \right]$, which rotates the ideal circle by $e^{j\phi}$. With this definition, we obtain the following scattering coefficients in the presence of a small circuit asymmetry
 \begin{align}
	S_{11} \approx& -\frac{e^{j\phi} Q_{\rm l}'/|Q_{\rm c}'|}{1+j2Q_{\rm l}'\delta}, \label{eq:asy_hanger4_s11}\\
	S_{21} = S_{12} \approx& 1 -\frac{e^{j\phi} Q_{\rm l}'/|Q_{\rm c}'|}{1+j2Q_{\rm l}'\delta}, \label{eq:asy_hanger4_s21}\\
	S_{22} \approx& -\frac{e^{j\phi} Q_{\rm l}'/|Q_{\rm c}'|}{1+j2Q_{\rm l}'\delta}. \label{eq:asy_hanger4_s22}
\end{align}
In what follows, we denote the expression of $Q'_{\rm l}$ and ${\rm Re}\left(Q_{\rm c}'\right)$ as the formal definition of loaded and coupling quality factors, and do not distinguish their notations from $Q_{\rm l}$ and $Q_{\rm c}$.

\subsubsection{Necklace- and bridge-type resonators}
For the necklace-type resonator shown in Fig.\,\ref{fig:load}(b), the elements of the transfer matrix with asymmetry can be written as 
\begin{align}
	A =& A_0 + \sinh \gamma l \left(\frac{\Delta{Z_1}}{{Z_0}}\right), \nonumber \\
	B =& B_0 
	+ \sinh \gamma l \left(\frac{\Delta{Z_1}\Delta{Z_2}}{Z_0} + \frac{\Delta{Z_1}}{j\omega C_2 Z_0} + \frac{\Delta{Z_2}}{j\omega C_1 Z_0}\right) \nonumber \\
	&+ \cosh \gamma l\left(\Delta{Z_1} + \Delta{Z_2}\right), \nonumber \\
	C =& C_0, \nonumber \\
	D =& D_0 + \sinh \gamma l \left(\frac{\Delta{Z_2}}{{Z_0}}\right), \nonumber 
\end{align}
where we denote the transfer matrix elements of the symmetric necklace-type resonator as $A_0$, $B_0$, $C_0$, and $D_0$. Following the derivation in Appendix\,\ref{app:asy_necklace2} and using the conventional definition $1/Q_{\rm l}'=1/Q_{\rm i} + {\rm Re}\left(1/Q_{\rm c}'\right)$, we obtain the scattering coefficients as
\begin{align}
	S_{11} \approx& 1 - \frac{e^{j\phi_1} 2Q_{\rm l}'/\left|Q_{{\rm c},1}'\right|}
	{1 + j2Q_{\rm l}'\delta}, \label{eq:asy_necklace2_s11}\\
	S_{21} = S_{12} \approx& \frac{e^{j\phi} 2Q_{\rm l}'/\sqrt{\left|Q_{{\rm c},1}'\right|\left|Q_{{\rm c},2}'\right|}}
	{1 + j2Q_{\rm l}'\delta}, \label{eq:asy_necklace2_s21}\\
	S_{22} \approx& 1 - \frac{e^{j\phi_2} 2Q_{\rm l}'/\left|Q_{{\rm c},2}'\right|}
	{1 + j2Q_{\rm l}'\delta}. \label{eq:asy_necklace2_s22}
\end{align}
Here, the coupling quality factor is defined as $1/Q_{\rm c}'=1/Q_{{\rm c},1}'+1/Q_{{\rm c},2}'$ with $Q_{{\rm c},1}' = Q_{{\rm c},1}Z_0/Z_1$ and $Q_{{\rm c},2}' = Q_{{\rm c},2}Z_0/Z_2$. The corresponding phases are $\phi = \left(\phi_1+\phi_2\right)/2$, $\phi_1 = -\arctan \left[ {\rm Im}\left( Q_{{\rm c},1}'\right)/{\rm Re} \left(Q_{{\rm c},1}' \right) \right]$, and $\phi_2 = -\arctan \left[ {\rm Im}\left( Q_{{\rm c},2}'\right)/{\rm Re} \left(Q_{{\rm c},2}' \right) \right]$.

The above results apply also to a bridge-type resonator, as shown in Fig.\,\ref{fig:load}(c), but with a $\pi$ phase difference in the transmission coefficients. The scattering coefficients read (see Appendix\,\ref{app:asy_cross2} for details) 
\begin{align}
	S_{11} \approx& 1 - \frac{e^{j\phi_1} 2Q_{\rm l}'/\left|Q_{{\rm c},1}'\right|}
	{1 + j2Q_{\rm l}'\delta}, \label{eq:asy_cross2_s11}\\
	S_{21} = S_{12} \approx& - \frac{e^{j\phi} 2Q_{\rm l}'/\sqrt{\left|Q_{{\rm c},1}'\right|\left|Q_{{\rm c},2}'\right|}}
	{1 + j2Q_{\rm l}'\delta}, \label{eq:asy_cross2_s21},\\
	S_{22} \approx& 1 - \frac{e^{j\phi_2} 2Q_{\rm l}'/\left|Q_{{\rm c},2}'\right|}
	{1 + j2Q_{\rm l}'\delta}. \label{eq:asy_cross2_s22}
\end{align}

In these regards, we conclude that a small circuit asymmetry leads to a small rotation of the ideal scattering-response circle in the complex plane. For the transmission coefficients of a hanger-type resonator or the reflection coefficients of a necklace- or bridge-type resonator, this rotation is centered at the reference point $\left(1+j0\right)$ with an angle $\phi$. This property can be used to correct for distortions caused by the circuit asymmetry in the measured scattering coefficients. 

In common experiments of superconducting quantum circuits, $\phi$ is often a small value such that our assumption of a small circuit asymmetry is valid. According to the literature, a huge circuit asymmetry is expected when the circuit shows a huge impedance mismatch that leads to multiple reflections in the cable \cite{Khalil2012, Deng2013}. However, we note that the above analyses may still apply and one can use the phase parameter $\phi$ to correct the circuit asymmetry from the measured scattering coefficients, as reported in Refs.\,\onlinecite{Petersan1998, Khalil2012, Deng2013, Probst2014, McRae2020}.

\subsection{Influence of the finite-length feedlines}\label{sec:finite}
Besides the circuit asymmetry, a finite length of the microwave feedlines can also influence the measured scattering coefficients, as shown in Fig.\,\ref{fig:load}(a)-(c). We recall the expression of the incident and reflected voltages at the position $z$ of a transmission line as \cite{Pozar2011}
\begin{equation}
	V^{\pm}(z) = V^{\pm}_0 e^{\mp\gamma z},
\end{equation}
where the $\pm$ sign distinguishes the incident and reflected wave propagations, and $V^{\pm}_0$ are the corresponding voltage amplitudes at $z=0$. For finite lengths of $l_1$ and $l_2$, the incident and reflected voltages transferred through the feedlines, $V_1^{\pm}{}'$ and $V_2^{\pm}{}'$, can be described as
\begin{equation}
	V_1^{\pm}{}' = e^{\pm \gamma l_1} V_1^{\pm},\,
	V_2^{\pm}{}' = e^{\pm \gamma l_2} V_2^{\pm}.
\end{equation}
Here, we denote $V_1^{\pm}$ and $V_2^{\pm}$ as the voltage amplitudes at the sample input and output. The scattering coefficients measured through the feedlines are thus
\begin{align}
	S_{11}' = e^{-2\gamma_1 l_1}S_{11}, \\
	S_{21}' = e^{-\left( \gamma_1 l_1 + \gamma_2 l_2 \right)}S_{21}, \\
	S_{12}' = e^{-\left( \gamma_1 l_1 + \gamma_2 l_2 \right)}S_{12}, \\
	S_{22}' = e^{-2\gamma_2 l_2}S_{22},
\end{align}
where we have assumed a perfect impedance match between the feedlines and the sample. A mismatched feedline causes a circuit asymmetry, which has already been discussed in the previous section. 

Under the high-frequency and low-loss approximations of the transmission feedlines, $\alpha$ is a constant and $\beta = \omega v_{\rm p}$ \cite{Pozar2011}. This reveals that the finite-length feedlines can cause a damping coefficient $A$ and a frequency-dependent phase factor, $e^{-j\omega\tau}$, in the scattering coefficients. Here, $\tau$ is a constant. In addition, there may also exist a constant phase delay, $e^{-j\varphi}$, because of the imperfect calibration of the cable delay. It can also be attributed to the circuit asymmetry, where a global phase factor was neglected in the previous section under the small asymmetry assumption. 

In total, we obtain a general model that describes the transmission coefficient of a hanger-type resonator or the reflection coefficient of a necklace- or bridge-type resonator 
\begin{equation}
	S\left(\omega\right) \approx A e^{-j\left(\omega\tau + \varphi\right)} \left( 1 - \frac{e^{j\phi}Q_{\rm l}/Q_{\rm c}}{1+2jQ_{\rm l}\left( \omega/\omega_{\rm r} -1 \right)} \right). \label{eq:s}
\end{equation}
This formula describes the scattering coefficients of a single resonator that can be measured in real experiments. Here, the global factor $A e^{-j\left(\omega\tau + \varphi\right)}$ originates from the finite length of the feedlines, while the local phase $e^{j\phi}$ results from a small circuit asymmetry. In the complex plane, the former rotates the circle of the ideal scattering coefficientsaround the original point $(0+j0)$, which is accompanied with a shrink of the circle radius. The latter causes a rotation around the reference point $(1+j0)$. 

As a closing remark, we note that Eq.\,\eqref{eq:s} applies only to common circuit geometries as discussed above. In a more complex circuit, for example, when two necklace-type $\lambda/2$ resonators are coupler in a chain \cite{Fischer2021}, Eq.\,\eqref{eq:s} cannot directly apply. Nevertheless, one can follow the transfer-matrix method to calculate the exact scattering coefficients of an arbitrary resonator network. Then, the influences of circuit asymmetry and feedlines can be added by hand. We will show in a parallel paper that a quantum-mechanical perspective of the scattering coefficients provides a shortcut to this calculation \cite{Chen2021b}.

\section{Experimental Results} \label{sec:experiment}
\begin{figure}
\centering
\includegraphics[width=\columnwidth]{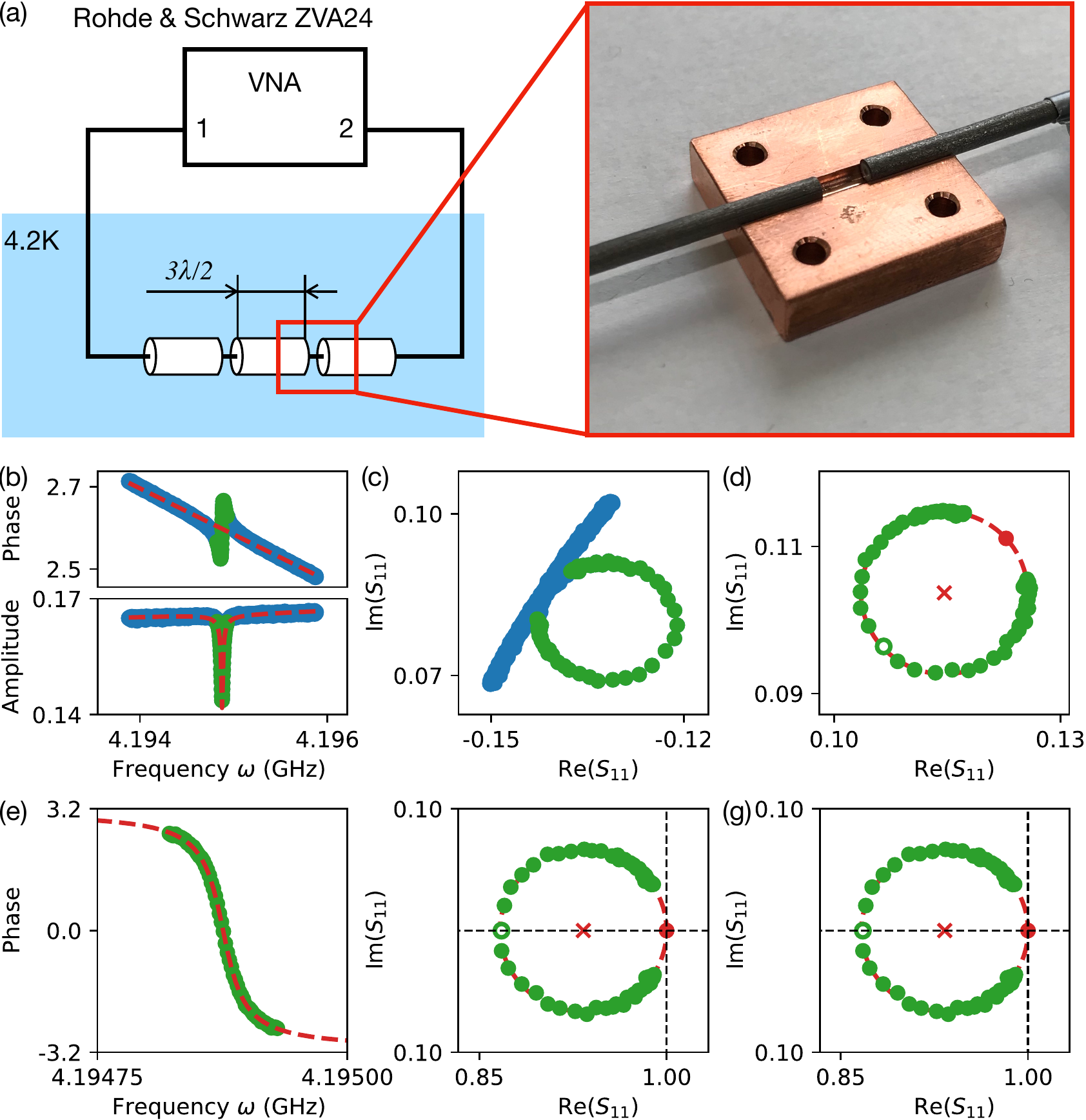}
\caption{(a) Sketch of the experimental setup used to measure the scattering coefficients of the necklace-type resonator with two mechanically tunable coupling capacitances by a vector network analyzer (VNA). (b) The step by step correction procedure for the measured reflection coefficient $S_{11}$ of the third harmonic mode of the necklace-type resonator shown in (a). Panels (b)-(g) correspond to the correction procedures described in Appendix\,\ref{app:circle-fit}.}
\label{fig:experiment}
\end{figure}

With the understandings of the line shape in measured scattering coefficients, we perform a prototypical experiment to characterize a necklace-type resonator by measuring only the reflection coefficient. The experimental setup is schematically shown in Fig.\,\ref{fig:experiment}(a). Here, the resonator consists of a commercial NbTi transmission line with length $l\simeq 81\,{\rm mm}$, which is capacitively coupled to two feedlines through small gaps with a customizable distance $d$. In our experiment, we focus on the third harmonic mode of the resonator, and measure the reflection coefficient, $S_{11}$, of the resonator with a fixed gap $d\simeq 1.5\,{\rm mm}$. The raw data, i.e., the amplitude $\left|S_{11}\right|$, phase $\arg\left[ S_{11}(\omega) \right]$, and the complex value $S_{11}(\omega)$, are shown in the blue dots in Fig.\,\ref{fig:experiment}(b)-(c), correspondingly. Following the recipe described in Appendix\,\ref{app:circle-fit}, we correct the influence of the finite feedlines and the circuit asymmetry sequentially, as shown in Fig.\,\ref{fig:experiment}(d)-(g). We obtain the resonant frequency and the internal, coupling, and loaded quality factors as $\omega_{\rm 3r}=2\pi\times 4.195\,{\rm GHz}$, $Q_{\rm 3l}=133946$, $Q_{\rm 3i}=154559$, and $Q_{\rm 3c}=1004318$. Here, the quality factors $Q_{\rm 3l}$, $Q_{\rm 3i}$, $Q_{\rm 3c}$ are defined for the third harmonic mode, which are different from those of the fundamental mode. The loaded quality factor for the $n$th mode is defined as $1/Q_{\rm nl}=1/Q_{\rm ni}+1/Q_{\rm nc}$, with the internal and coupling quality factors being, ideally,
\begin{align}
	Q_{\rm ni} =& \frac{n\pi}{2\alpha l} = nQ_{\rm i}, \\
	Q_{\rm nc} =& \frac{n\pi}{2\omega_{\rm nr}^2 Z_0^2 \left(C_1^2+C_2^2\right)} = \frac{Q_{\rm c}}{n}. 
\end{align}
This result is consistent with that obtained by the conventional method, which combines the measurement of both transmission and reflection coefficients. We note also that this method is also used in another experiment with two coupled necklace-type resonators \cite{Fischer2021}.

\section{Conclusions and outlook} \label{sec:conclusions}
In conclusion, we provide a comprehensive study of the scattering coefficients of superconducting microwave resonators in the classical perspective. By using the transfer-matrix method, we derive the analytical expressions of the scattering coefficients with different circuit geometries, such as hanger-, necklace-, and bridge-type. We also discuss the physical origin of distortions in the measured scattering coefficients, which include a finite circuit asymmetry and the effect of the finite length of the feedlines. These understandings open a door to correct the experimental imperfections in measurement results and characterize the electrical properties of a general microwave resonator. It is shown that, similar to the transmission coefficient of a hanger-type resonator, the reflection coefficient of a necklace- or bridge-type resonator contains a reference point which can be used to characterize the electrical properties of the resonator in a single measurement. We demonstrate this observation in experiment, and observe an excellent fit among the analytical, numerical, and experimental results.

\begin{acknowledgments}
We acknowledge support by German Research Foundation via Germany's Excellence Strategy (EXC-2111-390814868), Elite Network of Bavaria through the program ExQM, European Union via the Quantum Flagship project QMiCS (No.\,820505), German Federal Ministry of Education and Research via the project QuaRaTe (No.\,13N15380). The Python code for correcting experimental imperfections in measured scattering coefficients is available from the corresponding author on reasonable request, which will be open sourced shortly.
\end{acknowledgments}

\appendix
\begin{widetext}
\section{scattering coefficients of a necklace-type $\lambda/2$ resonator} \label{app:necklace2}
As discussed in Sec.\,\ref{sec:loaded_necklace2}, the total transfer matrix of a necklace-type $\lambda/2$ resonator is 
\begin{align}
	A =& \cosh \gamma l + \frac{\sinh \gamma l}{j\omega C_1 Z_0},&\,
	B =& \sinh \gamma l\left(Z_0 - \frac{1}{\omega^2C_1C_2Z_0} \right) + \cosh \gamma l \left(\frac{1}{j\omega C_1}+\frac{1}{j\omega C_2}\right), \\
	C =& \frac{\sinh \gamma l}{Z_0},&\,
	D =& \cosh \gamma l + \frac{\sinh \gamma l}{j\omega C_2 Z_0}.
\end{align}
Correspondingly, we write the elements of the corresponding scattering coefficients as
\begin{align}
	S_{11} =& \frac{\left( 2j\omega Z_0 C_2 +1 \right)e^{\gamma l}+\left( 2j\omega Z_0 C_1 -1 \right)e^{-\gamma l}}{\left( 2j\omega Z_0C_1 + 1 \right)\left(2j\omega Z_0C_2 + 1 \right)e^{\gamma l} - e^{-\gamma l}},\\
	S_{21} =& \frac{-4\omega^2Z_0^2C_1C_2}{\left( 2j\omega Z_0C_1 + 1 \right)\left(2j\omega Z_0C_2 + 1 \right)e^{\gamma l} - e^{-\gamma l}},\\
	S_{22} =& \frac{\left( 2j\omega Z_0 C_1 +1 \right)e^{\gamma l}+\left( 2j\omega Z_0 C_2 -1 \right)e^{-\gamma l}}{\left( 2j\omega Z_0C_1 + 1 \right)\left(2j\omega Z_0C_2 + 1 \right)e^{\gamma l} - e^{-\gamma l}},
\end{align}
where $S_{12}=S_{21}$. For open-circuited $\lambda/2$ resonators, we define $\beta l = \pi + \pi \Delta/\omega_0$ and $\alpha l=\pi/\left( 2Q_{\rm i} \right)$, and assume that the frequency $\omega$ is very close to the bare resonant frequency $\omega_0$. The above equations can be simplified as
\begin{align}
	S_{11} \approx& \frac{\left(\frac{1}{Q_{\rm i}} - \frac{2\omega Z_0\left(C_2 - C_1\right)\Delta}{\omega_0} \right) + j\left(\frac{2\omega Z_0\left( C_1 + C_2 \right)}{\pi} + \frac{2\Delta}{\omega_0} \right)}
	{\left(\frac{1}{Q_{\rm i}} - \frac{4\omega^2Z_0^2C_1C_2}{\pi} - \frac{2\omega Z_0 \left(C_1+C_2\right)\Delta}{\omega_0} \right) + j\left( \frac{2\omega Z_0\left( C_1 + C_2 \right)}{\pi} + \frac{2\Delta}{\omega_0} \right)},\\
	S_{21} \approx& \frac{\frac{4\omega^2Z_0^2C_1C_2}{\pi}}
	{\left(\frac{1}{Q_{\rm i}} - \frac{4\omega^2Z_0^2C_1C_2}{\pi} - \frac{2\omega Z_0 \left(C_1+C_2\right)\Delta}{\omega_0} \right) + j\left( \frac{2\omega Z_0\left( C_1 + C_2 \right)}{\pi} + \frac{2\Delta}{\omega_0} \right)},\\
	S_{22} \approx& \frac{\left(\frac{1}{Q_{\rm i}} - \frac{2\omega Z_0\left(C_1 - C_2\right)\Delta}{\omega_0} \right) + j\left(\frac{2\omega Z_0\left( C_1 + C_2 \right)}{\pi} +  \frac{2\Delta}{\omega_0} \right)}
	{\left(\frac{1}{Q_{\rm i}} - \frac{4\omega^2Z_0^2C_1C_2}{\pi} - \frac{2\omega Z_0 \left(C_1+C_2\right)\Delta}{\omega_0} \right) + j\left( \frac{2\omega Z_0\left( C_1 + C_2 \right)}{\pi} + \frac{2\Delta}{\omega_0} \right)}.
\end{align}
Similar to the analysis in the hanger-type resonator, the coupled resonant frequency $\omega_{\rm r}$ is obtained by setting the imaginary part of the denominator to \textit{zero}. This gives
\begin{equation}
	\omega_{\rm r} \approx \omega_0 - \omega_0^2 Z_0\left( C_1 + C_2 \right)/\pi,
\end{equation}
where we have assumed that $\omega_{\rm r}$ is close to $\omega_0$. Inserting this relation into the above equations, we obtain
\begin{align}
	S_{11} \approx& \frac{\frac{1}{Q_{\rm i}} + \frac{2\omega^2 Z_0^2\left(C_2^2 - C_1^2\right)}{\pi}+ j2\delta}
	{\left(\frac{1}{Q_{\rm i}} + \frac{2\omega^2Z_0^2 \left(C_1^2+C_2^2\right)}{\pi} \right) + j2\delta},\\
	S_{21} \approx& \frac{4\omega^2Z_0^2C_1C_2}
	{\left(\frac{1}{Q_{\rm i}} + \frac{2\omega^2Z_0^2 \left(C_1^2+C_2^2\right)}{\pi} \right) + j2\delta},\\
	S_{22} \approx& \frac{\frac{1}{Q_{\rm i}} + \frac{2\omega^2 Z_0^2\left(C_1^2 - C_2^2\right)}{\pi}+ j2\delta}
	{\left(\frac{1}{Q_{\rm i}} + \frac{2\omega^2Z_0^2 \left(C_1^2+C_2^2\right)}{\pi} \right) + j2\delta}.
\end{align}
Here, we have used the assumption that $\omega \approx \omega_{\rm r}$. Using again the assumption that $\omega$, $\omega_0$, and $\omega_{\rm r}$ are close to each other, and replace $\omega$ in the numerator by $\omega_{0}$, we obtain Eqs.\,\eqref{eq:necklace2_s11}-\eqref{eq:necklace2_s22} in the main text.

\section{scattering coefficients of a bridge-type $\lambda/2$ resonator} \label{app:cross2}
As discussed in Sec.\,\ref{sec:loaded_cross2}, the total transfer matrix of a bridge-type $\lambda/2$ resonator is  
\begin{align}
	A =& 1 + \frac{2\left( \alpha \lambda/4 + j\pi \Delta/2\omega_0 \right)}{j\omega C_1 Z_0},&\,
	B =&-\frac{2\left( \alpha \lambda/4 + j\pi \Delta/2\omega_0 \right)}{\omega^2C_1C_2Z_0}
	+ \left(\frac{1}{j\omega C_1}+\frac{1}{j\omega C_2}\right), \\
	C =& \frac{2\left( \alpha \lambda/4 + j\pi \Delta/2\omega_0 \right)}{Z_0},&\,
	D =& 1 + \frac{2\left( \alpha \lambda/4 + j\pi \Delta/2\omega_0 \right)}{j\omega C_2 Z_0}.
\end{align}
Correspondingly, we write the elements of the corresponding scattering coefficients as
\begin{align}
	S_{11} =& \frac{j\omega (C_1+C_2) Z_0 
	+2\left(\alpha \lambda/4 + j\pi \Delta/2\omega_0 \right)
	\left[1 + j\omega (C_2-C_1) Z_0 
	+ \omega^2C_1C_2Z_0^2\right]}
	{\left[-2\omega^2C_1C_2Z_0^2 
	+ j\omega (C_1+C_2) Z_0 \right]
	+2\left(\alpha \lambda/4 + j\pi \Delta/2\omega_0 \right)
	\left[1 + j\omega (C_1+C_2) Z_0 
	- \omega^2C_1C_2Z_0^2\right]},\\
	S_{21} =& \frac{-2\omega^2C_1C_2Z_0^2}
	{\left[-2\omega^2C_1C_2Z_0^2 
	+ j\omega (C_1+C_2) Z_0 \right]
	+2\left(\alpha \lambda/4 + j\pi \Delta/2\omega_0 \right)
	\left[1 + j\omega (C_1+C_2) Z_0 
	- \omega^2C_1C_2Z_0^2\right]},\\
	S_{22} =& \frac{j\omega (C_1+C_2) Z_0 
	+2\left(\alpha \lambda/4 + j\pi \Delta/2\omega_0 \right)
	\left[1 + j\omega (C_1-C_2) Z_0 
	+ \omega^2C_1C_2Z_0^2\right]}
	{\left[-2\omega^2C_1C_2Z_0^2 
	+ j\omega (C_1+C_2) Z_0 \right]
	+2\left(\alpha \lambda/4 + j\pi \Delta/2\omega_0 \right)
	\left[1 + j\omega (C_1+C_2) Z_0 
	- \omega^2C_1C_2Z_0^2\right]},
\end{align}
Following a similar procedure in Appendix.\,\ref{app:necklace2}, we omit the intermediate steps and obtain the following scattering coefficients
\begin{align}
	S_{11} \approx& \frac{\frac{1}{Q_{\rm i}} + \frac{2\omega (C_2^2-C_1^2) Z_0}{\pi}}{\left[\frac{1}{Q_{\rm i}}+ \frac{\omega_0^2 (C_1^2+C_2^2) Z_0^2}{\pi} \right]
	+ j2\delta},\\
	S_{21} \approx& \frac{\frac{-4\omega^2C_1C_2Z_0^2}{\pi}}
	{\left[\frac{1}{Q_{\rm i}}+ \frac{\omega_0^2 (C_1^2+C_2^2) Z_0^2}{\pi} \right]
	+ j2\delta},\\
	S_{22} \approx& \frac{\frac{1}{Q_{\rm i}} + \frac{2\omega (C_1^2-C_2^2) Z_0}{\pi}}{\left[\frac{1}{Q_{\rm i}}+ \frac{\omega_0^2 (C_1^2+C_2^2) Z_0^2}{\pi} \right]
	+ j2\delta}.
\end{align}
Here, the resonance frequency is
\begin{equation}
	\omega_{\rm r} = \omega_0 - Z_0\omega_0^2\left( C_1 + C_2 \right)/\pi.
\end{equation} 

\section{Asymmetry in hanger-type $\lambda/4$ resonator} \label{app:asy_hanger4}
With circuit asymmetries $\Delta{Z_1}$ and $\Delta{Z_2}$, we write the scattering coefficients as
\begin{align}
	S_{11} =& \frac{\left( \Delta{z_1} + \Delta{z_2} \right)z_3 + \left(\Delta{z_1} -1\right)z_2}{\left( z_1 + z_2 \right)z_3+ z_1z_2},\\
	S_{21} =& S_{12} = \frac{2z_3}{\left( z_1 + z_2 \right)z_3+ z_1z_2},\\
	S_{22} =& \frac{\left( \Delta{z_1} + \Delta{z_2} \right)z_3 + \left(\Delta{z_2} -1\right)z_1}{\left( z_1 + z_2 \right)z_3+ z_1z_2}.
\end{align}
Here, $\Delta{z_1} = \Delta{Z_1}/Z_0$, $\Delta{Z_2} = \Delta{Z_2}/Z_0$. To proceed, we define the coupling quality factor $Q_{\rm c}' = Q_{\rm c}\left[ 1/(2z_1) + 1/(2z_2) \right]$. The above expression can be simplified as
\begin{align}
	S_{11} \approx& \left(1- \frac{2}{z_1+z_2}\right) - \frac{2z_2}{\left(z_1+z_2\right)z_1}\frac{Q_{\rm l}'/Q_{\rm c}'}{1 + 2Q_{\rm l}'\delta},\\
	S_{21} =& S_{12} \approx \frac{2}{z_1+z_2}\left(1-\frac{Q_{\rm l}'/Q_{\rm c}'}{1 + 2Q_{\rm l}'\delta}\right),\\
	S_{22} \approx& \left(1- \frac{2}{z_1+z_2}\right) - \frac{2z_1}{\left(z_1+z_2\right)z_2}\frac{Q_{\rm l}'/Q_{\rm c}'}{1 + 2Q_{\rm l}'\delta},
\end{align}
where $z_1 = 1 + \Delta{z_1}$, $z_2 = 1 + \Delta{z_2}$. For small asymmetries $z_1 \approx z_2 \approx 1$, we obtain the final form of the scattering coefficients described in Eqs.\,\eqref{eq:asy_hanger4_s11}-\eqref{eq:asy_hanger4_s22}.

\section{Asymmetry in necklace-type $\lambda/2$ resonator} \label{app:asy_necklace2}
With circuit asymmetries $\Delta{Z_1}$ and $\Delta{Z_2}$, we write the scattering coefficients as
{\small\begin{align}
	S_{11} \approx& \frac{\left[-\omega^2c_1c_2\left(z_1 + z_2 -2 \right) + j\omega \left(c_1 + c_2\right) \right] 
	+ \left( \alpha l + j\beta l \right) \left[-\omega^2c_1c_2z_1z_2 + j\omega \left(c_1z_1 + c_2z_2\right) +1 - 2\omega^2c_1c_2\left( 1 - z_2\right) - 2j\omega c_1 \right]}
	{\left[-\left(z_1 + z_2 \right)\omega^2 c_1 c_2 + j\omega \left( c_1 + c_2 \right) \right] + \left( \alpha l + j\beta l \right) \left( -z_1z_2\omega^2 c_1 c_2 + j\omega \left(z_1c_1 + z_2c_2 \right) + 1 - \omega^2 c_1 c_2 \right)},\\
	S_{21} =& S_{12} \approx \frac{2\omega^2 c_1 c_2}
	{\left[-\left(z_1 + z_2 \right)\omega^2 c_1 c_2 + j\omega \left( c_1 + c_2 \right) \right] + \left( \alpha l + j\beta l \right) \left( -z_1z_2\omega^2 c_1 c_2 + j\omega \left(z_1c_1 + z_2c_2 \right) + 1 - \omega^2 c_1 c_2 \right)},\\
	S_{22} \approx& \frac{\left[-\omega^2c_1c_2\left(z_1 + z_2 -2 \right) + j\omega \left(c_1 + c_2\right) \right] + \left( \alpha l + j\beta l \right) \left[-\omega^2c_1c_2z_1z_2 + j\omega \left(c_1z_1 + c_2z_2\right) +1 - 2\omega^2c_1c_2\left( 1 - z_1\right) - 2j\omega c_2 \right]}
	{\left[-\left(z_1 + z_2 \right)\omega^2 c_1 c_2 + j\omega \left( c_1 + c_2 \right) \right] + \left( \alpha l + j\beta l \right) \left( -z_1z_2\omega^2 c_1 c_2 + j\omega \left(z_1c_1 + z_2c_2 \right) + 1 - \omega^2 c_1 c_2 \right)}.
\end{align}}%
Here, we have replaced $\cosh \gamma l$ and $\sinh \gamma l$ by $-1$ and $-\left(\alpha l + j\beta l\right)$, respectively, under  the assumption of small loss and defined $c_1=Z_0 C_{1}$, $c_2=Z_0 C_{2}$. For small asymmetries $z_1 \approx z_2 \approx 1$ and small coupling capacitances $j\omega Z_0 C_1, \omega Z_0 C_2 \ll 1$, we simplify the expression as
\begin{align}
	S_{11} \approx& \frac{\frac{\pi}{2Q_{\rm i}} + \omega^2\left( z_2c_2^2 + c_1^2\left(z_1-2\right) \right) + j\pi\delta}
	{\left[\frac{\pi}{2Q_{\rm i}} + \omega^2 \left(z_1c_1^2 + z_2c_2^2 \right) \right]  + j\pi\delta},\\
	S_{21} =& S_{12} \approx \frac{2\omega^2 c_1 c_2}
	{\left[\frac{\pi}{2Q_{\rm i}} + \omega^2 \left(z_1c_1^2 + z_2c_2^2 \right) \right]  + j\pi\delta},\\
	S_{22} \approx& \frac{\frac{\pi}{2Q_{\rm i}} + \omega^2\left( z_1c_1^2 + c_2^2\left(z_2-2\right) \right) + j\pi\delta}
	{\left[\frac{\pi}{2Q_{\rm i}} + \omega^2 \left(z_1c_1^2 + z_2c_2^2 \right) \right]  + j\pi\delta}.
\end{align}
To proceed, we define $1/Q_{\rm c}' = 1/Q_{{\rm c},1}' + 1/Q_{{\rm c},2}'$ with $Q_{{\rm c},1}' = \pi/2\omega^2c_1^2z_1$, $Q_{{\rm c},2}' = \pi/2\omega^2c_2^2z_2$. The above expression can be simplified as
\begin{align}
	S_{11} \approx& 1 - \frac{1}{z_1}\frac{2Q_{\rm l}'/Q_{{\rm c},1}'}
	{1 + j2Q_{\rm l}'\delta}, \label{eq:asy_necklace2_s11_app}\\
	S_{21} = S_{12} \approx& \frac{1}{\sqrt{z_1z_2}} \frac{2Q_{\rm l}'/\sqrt{Q_{{\rm c},1}'Q_{{\rm c},2}'}}
	{1 + j2Q_{\rm l}'\delta},\\
	S_{22} \approx& 1 - \frac{1}{z_2}\frac{2Q_{\rm l}'/Q_{{\rm c},2}'}
	{1 + j2Q_{\rm l}'\delta}. \label{eq:asy_necklace2_s22_app}
\end{align}
For small asymmetries, we omit the phase contributions of $1/z_1$ and $1/z_2$, and obtain the final form of the scattering coefficients described in Eqs.\,\eqref{eq:asy_necklace2_s11}-\eqref{eq:asy_necklace2_s22}.

\section{Asymmetry in bridge-type $\lambda/2$ resonator} \label{app:asy_cross2}
With circuit asymmetries $\Delta{Z_1}$ and $\Delta{Z_2}$, we write the scattering coefficients as
{\small\begin{align}
	S_{11} =& \frac{\left[j\omega \left(c_1+c_2\right) - \omega^2c_1c_2\left(z_1 + z_2-2\right) \right] 
	+2\left(\alpha l + j\pi \Delta/2\omega_0 \right)\left[1+j\omega \left(c_1 z_1+c_2 z_2\right) - \omega^2 c_1c_2z_1z_2 + 2\omega^2c_1c_2z_2 - 2j\omega c_1\right]}
	{\left[j\omega \left(c_1+c_2\right) - \omega^2c_1c_2\left(z_1 + z_2\right) \right]
	+2\left(\alpha l + j\pi \Delta/2\omega_0 \right)
	\left[1+j\omega \left(c_1 z_1+c_2 z_2\right) - \omega^2 c_1c_2z_1z_2 \right]}, \\
	S_{21} =& S_{12} = \frac{-2\omega^2 c_1c_2}
	{\left[j\omega \left(c_1+c_2\right) - \omega^2c_1c_2\left(z_1 + z_2\right) \right]
	+2\left(\alpha l + j\pi \Delta/2\omega_0 \right)
	\left[1+j\omega \left(c_1 z_1+c_2 z_2\right) - \omega^2 c_1c_2z_1z_2 \right]}, \\
	S_{22} =& \frac{\left[j\omega \left(c_1+c_2\right) - \omega^2c_1c_2\left(z_1 + z_2-2\right) \right] 
	+2\left(\alpha l + j\pi \Delta/2\omega_0 \right)\left[1+j\omega \left(c_1 z_1+c_2 z_2\right) - \omega^2 c_1c_2z_1z_2 + 2\omega^2c_1c_2z_1 - 2j\omega c_2\right]}
	{\left[j\omega \left(c_1+c_2\right) - \omega^2c_1c_2\left(z_1 + z_2\right) \right]
	+2\left(\alpha l + j\pi \Delta/2\omega_0 \right)
	\left[1+j\omega \left(c_1 z_1+c_2 z_2\right) - \omega^2 c_1c_2z_1z_2 \right]}.
\end{align}}%
For small coupling capacitances $j\omega Z_0 C_1, \omega Z_0 C_2 \ll 1$, the above formulae are equivalent to those for a necklace-type $\lambda/2$ resonator, except the $\pi$ phase difference in the transmission coefficient. In this regard, the scattering coefficients with regards to small circuit asymmetries are also similar to Eqs.\,\eqref{eq:asy_necklace2_s11_app}-\eqref{eq:asy_necklace2_s22_app}. The complete results are shown in Eqs.\,\eqref{eq:asy_cross2_s11}-\eqref{eq:asy_cross2_s22} in the main text.

\section{Correction of experimental imperfections} \label{app:circle-fit}
\begin{figure}
\centering
\includegraphics[width=\columnwidth]{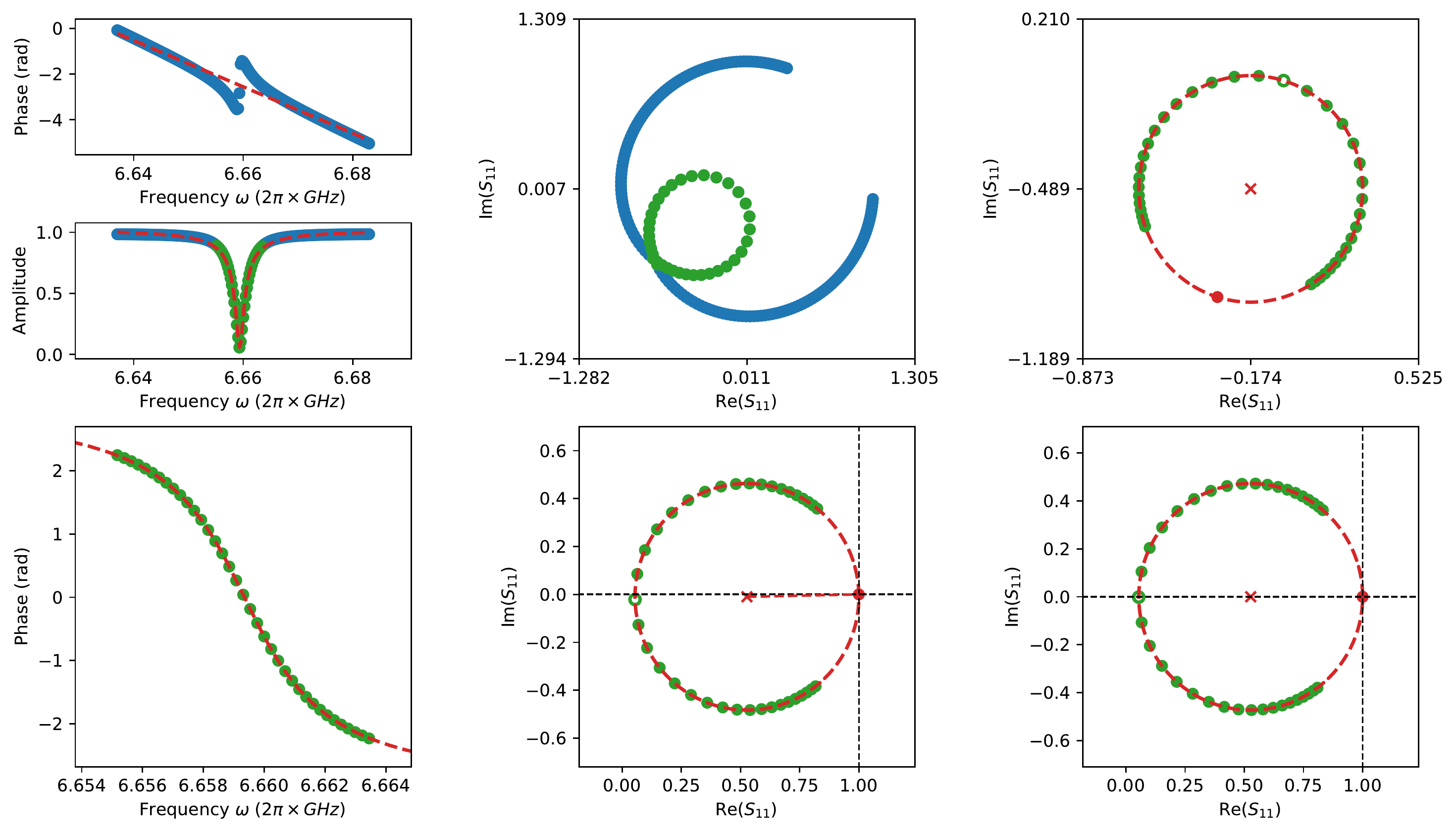}
\caption{The step by step correction procedure for the simulated reflection coefficient $S_{11}$ of a necklace-type microwave resonator. (a) Rough phase correction with linear fit, and data selection with Lorentzian fit, where the green dots are the data within the $\pm 4\Delta_{\rm 3dB}$ bandwidth. (b) The raw data and the selected data shown in the complex plane. (c) Fine phase correction with circle fit. (d) Resonant frequency determination with phase versus frequency fit. (e) Correction of cable attenuation and frequency-independent phase. (f) Correction of asymmetry. In all the plots, the blue dots denote raw data and the green dots the selected data. The fitting results are colored in red, where the red cross and dot represent the circle center and the far off-resonant point on the circle, respectively.}
\label{fig:correction}
\end{figure}

We now take the necklace-type resonator, shown in Fig.\,\ref{fig:load}(c), as an example and describe a circle-fitting procedure that corrects the experimental imperfections in the reflection coefficient, $S_{11}$. Here, the test data is generated by using a distributed-element circuit model, where the parameters can be fully controlled as a crosscheck of our results. We choose $\alpha=5.0\times 10^{-3}\,{\rm /m}$, $v_{\rm p}=1.35\times 10^{8}\,{\rm m/s}$, $C_1, C_2=1.0\times 10^{-14}\,{\rm F}$, and $l=1\times 10^{-2}\,{\rm m}$ for a necklace-type $\lambda/2$ resonator. We also assume a small circuit asymmetry, $\Delta{Z_1} = j\omega L_1$ with $L_1=1 \times 10^{-9}\,{\rm H}$, $\Delta{Z_2} = 2\,{\rm \Omega}$, and consider two finite-length feedlines with $l_1=l_2=1.2\,{\rm m}$ with a $50\,{\rm \Omega}$ impedance match. In these regards, one can simulate the reflection coefficient as shown in Fig.\,\ref{fig:correction}(a)-(b). The resonant frequency is estimated to be $\omega_{\rm r}=2\pi\times 6.659\,{\rm GHz}$, and the loaded, internal, and coupling quality factors are $Q_{\rm l}=1666$, $Q_{\rm i}=31416$, and $Q_{\rm c}=1759$, respectively. The correction procedure consists of three major steps:

\subsection{Correction of frequency-dependent phase shift}
In the first step, we eliminate the frequency-dependent phase shift $e^{-j\left(\omega\tau+\varphi_1\right)}$ with $\varphi_1$ being an arbitrary phase offset. This phase shift can be directly seen in Fig.\,\ref{fig:correction} (b), where an expected circle of the scattering coefficient is distorted into a knot in the complex plane. The elimination process can be implemented as follows: 

First, we use a linear function to fit $\tau$ and $\varphi_1$, which serve as an initial guess for a more precise fitting procedure. The objective function is
\begin{equation}
	J_1 = \left\{ -(\omega\tau + \varphi_1) - \arg\left[S\left(\omega\right)\right] \right\}^2,
\end{equation}
where $\arg\left[S\left(\omega\right)\right]$ is the unwrapped phase of the complex signal. The fitting result is shown as the red dashed line in Fig.\,\ref{fig:correction}(a), where $\tau=16.18\,{\rm s}$. 

Next, we use a Lorentzian function to determine the full width at half maximum (FWHM), $\Delta_{\rm 3dB}$, of the line shape, which is used to select data that are close to the resonant frequency. The objective function is \cite{Petersan1998} 
\begin{equation}
	J_2 = \left[ \left(A_1 + A_2f 
	+ \frac{A_3 + A_4f}{\sqrt{1+ 4\left( \frac{\omega - \omega_{\rm r}}{\Delta_{\rm 3dB}} \right)^2}}\right) - \left|S\left(\omega\right)\right| \right]^2.
\end{equation}
Here, $\omega_{\rm r}$, $\Delta_{\rm 3dB}$, and $A_1, \cdots,A_4$ are fitting parameters. Because the data within the FWHM can form a half circle in the ideal case, to minimize the influence of the background signal we keep only the data within $3$-$5$ times of the FWHM in the following analyses. The fitting result is shown as the green dots in Fig.\,\ref{fig:correction}(a)-(b), where $\omega_{\rm r}=2\pi\times 6.659\,{\rm GHz}$ and $\Delta_{\rm 3dB}=2\pi\times 1.41\,{\rm MHz}$.  

Having obtained the initial guess of the parameters $\tau$, $\varphi_1$ and removed the far off-resonant data points beyond $3\Delta_{\rm 3dB}$-$5\Delta_{\rm 3dB}$, we correct the frequency-dependent phase shift by using the circuit-fit technique \cite{Probst2014}. On the one hand, we use an algebraic method to fit a circle to the scattering coefficients\cite{Chernov2005}, and determine the circle center $S_{\rm c}$ and the radius $r_{\rm c}$. On the other hand, we can optimize the parameter $\tau$ with the following objective function
\begin{equation}
	J_3 = \left( r_{\rm c} - \left|e^{j\omega\tau}S\left( \omega \right) - S_{\rm c1}\right| \right)^2,
\end{equation}
to make the corrected data $S_1\left( \omega \right) = e^{j\omega\tau}S\left( \omega \right)$ more likely to be a circle. The corrected data $S_1\left( \omega \right)$ and the fitted circle is shown in Fig.\,\ref{fig:correction}(c), where $\tau=17.58\,{\rm s}$, $S_{\rm c1}=-0.218-j0.471$.

\subsection{Correction of attenuation and frequency-independent phase shift}
After correcting the phase shift $e^{-j\left(\omega\tau+\varphi_1\right)}$, the next step is to eliminate the attenuation and the frequency-independent phase shift $Ae^{-j\varphi_2}$ with $\varphi_2=\varphi-\varphi_1$. We recall the fact that, without the influence of the finite feedline, the reflection coefficient of a necklace-type resonator interacts with the real axis at the reference point $(1 + j0)$ at a far-detuned probe frequency $\omega \rightarrow \infty$. We use this property to correct the attenuation and frequency-independent phase shift.

First, we determine the resonant frequency, $\omega_{\rm r}$ by using a phase-versus-frequency fit, which is proven to be the most precise and robust fitting method for calibrating a microwave resonator \cite{Petersan1998}. The objective function is 
\begin{equation}
	J_4 = \left\{ \varphi_2 + 2\arctan\left[ 2Q_{\rm l}\left( 1- \frac{\omega}{\omega_{\rm r}} \right) \right] - \arg\left[S_1\left(\omega\right)\right] \right\}^2,
\end{equation}
where $\varphi_2=\left(\varphi-\varphi_1\right) + (2n+1)\pi$, $\omega_{\rm r}$, and $Q_{\rm l}$ are fitting parameters. The fitting results are shown in Fig.\,\ref{fig:correction}(d), from which we determine the resonant frequency $\omega_{\rm r}=2\pi\times 6.659\,{\rm GHz}$ and the loaded quality factor $Q_{\rm l}=1655$. 

Knowing the value of $\omega_{\rm r}$, one can locate the resonant data point in the fitted circle at $S_{\rm r}=-0.041-j0.040$. Correspondingly, the far off-resonant point, $S_{\rm off}=-0.395-j0.902$, is determined according to the symmetry of a circle, i.e., $S_{\rm off} = S_{\rm c1} + \left( S_{\rm c1} - S_{\rm r} \right)$. Then, one can correct the attenuation and the frequency-independent phase shift according to the following relation
\begin{equation}
	S_2(\omega) = S_1(\omega)/S_{\rm off}.
\end{equation}
The corrected data, $S_2(\omega)$, is shown in Fig.\,\ref{fig:correction}(e), where $S_{\rm r} =0.054-j0.021$, $S_{\rm off} =1.000$, $S_{\rm c2} = S_{\rm c1}/S_{\rm off}=0.527-j0.011$, $r_{\rm c2}=r_{\rm c1}/\left|S_{\rm off}\right|=0.473$. 

\subsection{Correction of circuit asymmetry}
After the first two steps, we have removed the influence of finite feedlines. The last step is to correct the circuit asymmetry. Here, we use the property that the circle center should be located on the real axis in an ideal scattering coefficient. We identify $\phi$ by the following relation
\begin{equation}
	\phi = \arg\left( S_{\rm c2} - S_{\rm off} \right) - \pi,
\end{equation}
and rotate the circle by $-\phi$ around the reference point $\left(1+j0\right)$. In the meantime, we also rescale the circle radius by a factor of $\left|\cos \phi\right|$ to account for the difference between $\left|Q_{\rm c}\right|$ and ${\rm Re}\left(Q_{\rm c}\right)$ \cite{Khalil2012}. In total, the transformation is described by
\begin{equation}
	S_3(\omega) = \cos\phi\left[S_2(\omega) - 1 \right]e^{-j\phi} + 1.
\end{equation}
The corrected reflection coefficient, $S_3(\omega)$, is shown in Fig.\,\ref{fig:correction}(f). In this example, we determine $\phi=0.023$ and thus $S_{\rm c3}=0.527$, $r_{\rm c3}=0.473$. The internal quality factor can be calculated as $Q_{\rm i} = Q_{\rm l}(1-2r_{\rm c3})=30530$, where $Q_{\rm c}=Q_{\rm l}/\left(2r_{\rm c3}\right)=1750$. The loaded quality factor $Q_{\rm l}=1655$ is obtained in the phase versus frequency fit. Comparing the fitted quality factors, $Q_{\rm l}$, $Q_{\rm i}$, and $Q_{\rm c}$, with the estimated values, we obtain the relative errors of the calibration results to be $0.6\%$, $2.8\%$, $0.5\%$, respectively. 

\end{widetext}

\bibliography{PT_ref}  
\end{document}